\documentclass[fleqn,10pt]{wlscirep}
\usepackage[utf8]{inputenc}
\usepackage[T1]{fontenc}
\usepackage{amssymb,amsthm,amsmath,bbm,bbold,graphicx,epsfig,threeparttable,color}
\usepackage{ulem,enumerate} 
\usepackage{enumitem}
\usepackage[title]{appendix}
\usepackage{hyperref}
\usepackage{lineno}

\newcommand{\ket}[1]{{|#1\rangle}}
\newcommand{\bra}[1]{{\langle#1|}}

\newcommand{\Tr}{{\rm Tr}}

\theoremstyle{definition}
\newtheorem{defi}{Definition}

\newtheorem{coro}[defi]{Corollary}

\title{Deep learning the hierarchy of steering measurement settings of qubit-pair states}

\author[1,2]{Hong-Ming Wang}
\author[3,*]{Huan-Yu Ku}
\author[1]{Jie-Yien Lin}
\author[1,2,4,$\dagger$]{Hong-Bin Chen}
\affil[1]{Department of Engineering Science, National Cheng Kung University, Tainan 701401, Taiwan}
\affil[2]{Physics Division, National Center for Theoretical Sciences, Taipei 10617, Taiwan}
\affil[3]{Department of Physics, National Taiwan Normal University, Taipei 116059, Taiwan}
\affil[4]{Center for Quantum Frontiers of Research \& Technology, NCKU, Tainan 701401, Taiwan}
\affil[*]{huan.yu@ntnu.edu.tw}
\affil[$\dagger$]{hongbinchen@gs.ncku.edu.tw}

\begin{abstract}
Quantum steering has attracted increasing research attention because of its fundamental importance, as well as its applications in quantum information science. Here we leverage the power of the deep learning model to infer the steerability of quantum states with specific numbers of measurement settings, which form a hierarchical structure. A computational protocol consisting of iterative tests is constructed to overcome the optimization, meanwhile, generating the necessary training data. According to the responses of the well-trained models to the different physics-driven features encoding the states to be recognized, we can numerically conclude that the most compact characterization of the Alice-to-Bob steerability is Alice's regularly aligned steering ellipsoid; whereas Bob's ellipsoid is irrelevant. We have also provided an explanation to this result with the one-way stochastic local operations and classical communication. Additionally, our approach is versatile in revealing further insights into the hierarchical structure of quantum steering and detecting the hidden steerability.
\end{abstract}

\begin{document}

\flushbottom
\maketitle

\section*{Introduction}

The quantum correlations between spatially separated parties are widely acknowledged as a nontrivial resource showing advantages over the classical counterparts for quantum foundations and quantum information~\cite{Arute2019,SchuldPRXQ2022,MonroeRMP2021,Daley2022}. Among the family of quantum correlations, Bell nonlocality \cite{brunner_bell_nonlocal_rmp_2014} is arguably the most famous one due to the historical debate of Einstein-Podolsky-Rosen paradox \cite{epr_paradox_pr_1935}. A nonlocal state can be certified by the violation of Bell's inequality \cite{bell_ineq_phys_1964}, and can be explicitly applied to the fully device-independent quantum information tasks \cite{acin_di_task_prl_2007,bancal_di_task_prl_2011,shin-liang_prl_2016,shin-liang_quantum_2021,shin-liang_prr_2021}.

Apart from the fully device-independent scenario, the constraints can be relaxed to the one-side device-independent quantum information tasks~\cite{Branciard2012,Piani2015,Skrzypczyk2018}, in which one of the two-party system (say Bob) is steered by the other party's (say Alice) untrusted measurements. In this Alice-to-Bob steering scenario, one relies on a steerable correlation, also referred to steerable assemblage, as a resource to achieve quantum advantages over an unsteerable assemblage admitting the local-hidden-state (LHS) models~\cite{Wiseman2007PRL,steering_rpp_2017,steering_rmp_2020}. Such a quantum resource requires a steerable state with an incompatible measurement~\cite{Quintino2014PRL,Uola2014PRL,Uola2015PRL} such that the generated assemblage is incompatible with the LHS models.

There have been many approaches to test the incompatibility of a given assemblage with respect to the LHS models~\cite{saunders_ste_ineq_np_2010,smith_ste_ineq_nc_2012,Zhao2020,KuPRXQ2022,Slussarenko2022-2,Zhao:23,Ku2023arXiv}; in contrast, to verify the steerability of a quantum state remains a cumbersome task. To that aim, one has to optimize over all possible incompatible measurements, e.g., (1) the number $n$ of observables and (2) which observables to be measured on a given state. With this construction, the classification of quantum states according to the minimal number $n$ of required observables to exhibit the steerability forms a hierarchy, which we refer to as the hierarchy of steering measurement settings.

There are some attempts to tackle this difficulty by deriving feasible criteria. For example, the criterion, derived by Bowles, Hirsch, Quintino, and Brunner (BHQB) \cite{bowles_bhqb_ste_cri_pra_2016}, is a sufficient condition for verifying the qubit-pair unsteerability. Furthermore, the maximal violation of the steering inequality, derived by Cavalcanti, Jones, Wiseman, and Reid (CJWR) \cite{PhysRevA.80.032112}, by a qubit-pair state under 2 and 3 observables has been utilized as a quantifier of qubit-pair steerablity \cite{PhysRevA.93.020103}. However, both conditions are not strong enough in fully discriminating the (un)steerability of qubit-pair states. The explicitly experimental examination of the insufficiency has been reported recently~\cite{Adam2021PRA}.

On the other hand, inspired by the wide-spreading applications of learning algorithms in quantum physics, e.g., quantum dynamics \cite{luchnikov_mach_lear_q_dyna_prl_2020,fanchini_mach_lear_q_dyna_pra_2021,goswami_mach_lear_q_dyna_pra_2021}, quantum computing
\cite{WisePRXQ2021,StrikisPRXQ2021}, quantum chemistry~\cite{garrido_mach_lear_appl_nc_2021,Xiao2022,Gebauer2022}, and quantum communication~\cite{WallnoferPRXQ2020,Chinnpj2021} (see also the recent reviews~\cite{carleo_mach_lear_appl_rmp_2019,karniadakis_mach_lear_appl_nrp_2021,krenn_mach_lear_appl_pra_2023}),
the machine-learning approach has also been applied to verify the quantum correlations~\cite{lu_mach_lear_sep_ent_par_2018,ma_mach_lear_bell_ineq_npjqi_2018,canabarro_mach_lear_nonlocality_prl_2019,
ren_mach_lear_steerability_pra_2019,krivachy_mach_lear_nonlocality_npjqi_2020,zhang_mach_lear_steerability_pra_2022,chen_mach_lear_multi_ent_qst_2023,girardin_mach_lear_sep_appr_prr_2022,
yang_mach_lear_multi_nonlocality_prl_2019,tian_mach_lear_multi_ent_prl_2022}. Although the idea of detecting the steerability of qubit-pair states with machine learning has been
implemented~\cite{ren_mach_lear_steerability_pra_2019,zhang_mach_lear_steerability_pra_2022}, their approach hardly reveals the physical insights, e.g., the minimal number of required observables to demonstrate steerability and characterization of the steerability of quantum states, from the prediction.

Here we leverage the power of the supervised deep learning algorithm to infer the fine-grained hierarchy of steering measurement settings. Due to the aforementioned insufficiency of the theoretical criteria, to exactly verify this hierarchy for quantum states is difficult, rendering the actual boundaries of the hierarchy undetermined. Such a problem without ground truth (GT), or hard to find out the ground truth, is called to be GT-deficient. This hinders the generation of reliable training data, as well as the training of a supervised deep learning algorithm.

To overcome this insufficiency, we construct a computational protocol by iteratively testing the steerability of a given qubit-pair state with the semidefinite program (SDP) under $n$ random observables, designating the state to be $n$-measurement steerable ($n$-MS). This constitutes the necessary data set for training a deep learning algorithm.

Physical intuition on the steerability allows us to reduce the parameters encoding a qubit-pair state in the data set. From the counterintuitive responses of the well-trained deep learning models to the physics-driven parameters, we can acquire a compact, and precise, way for the characterization of steerability. Our results suggest that, in contrast to the entanglement, the characterization of the steerability from Alice to Bob is dominantly determined by Alice's regularly aligned ellipsoid; whereas Bob's ellipsoid is irrelevant. Recall that Alice's steering ellipsoid~\cite{jevtic_steering_ellipsoid_prl_2014} is defined to be the set of all states that can be steered to from Bob, and vice versa. Comparing these operational definitions of quantum steering and the steering ellipsoid, our results are seemingly counterintuitive. Therefore, we have also provided an explanation with the one-way stochastic local operations and classical communication (1W-SLOCC).

For a comprehensive visualization, we have also applied the well-trained deep learning models to depict the hierarchy of two types of states generalized from the Werner state. Our models are capable of identifying the 4-MS states, i.e., the states requiring at least four observables to exhibit steerability on Bob, with high precision. Currently, no theoretical criteria in the literature are capable of efficiently determining the 4-MS states. Furthermore, with a different way of parameterizing the qubit-pair states, our models can also be applied to characterize hidden
quantum steerability of generalized Werner states, whereby an unsteerable state becomes steerable after an 1W-SLOCC operation. Our results, on the one hand, unravel a new direction in the characterization of steerability; on the other hand, these also underpin the capability of the deep learning approaches to shed light on a new route toward the physics remaining obscure, rather than merely being a substitution of cumbersome computational tasks.

\section*{Results}

\subsection*{Hierarchy of steering measurement settings}

We first recall the notion of quantum steering. Consider two parties, denoted as Alice and Bob, sharing an unknown quantum state $\rho^\mathrm{AB}$. Alice performs a measurement, described by the projective measurement $M_{a|x}$ satisfying $M_{a|x}M_{a'|x}=\delta_{a,a'}M_{a|x}~\forall~x$ and $\sum_aM_{a|x}=I_d$, where $x=0,\ldots,n-1$ and $a=0,\ldots,o-1$ represent the index of observables and outcomes of measurements, respectively. With one-way classical communication, Bob will obtain a set of (subnormalized) states, referred to as the assemblage $\{\sigma_{a|x}\}_{a,x}$, with $\sigma_{a|x}=\Tr_{\rm{A}}[(M_{a|x}\otimes I_d) \rho^\mathrm{AB}]$, containing both classical information of the probability $p(a|x)=\Tr(\sigma_{a|x})$ and Bob's quantum states $\rho^\mathrm{B}_{a|x}=\sigma_{a|x}/\Tr(\sigma_{a|x})$.

An assemblage is defined to be unsteerable if it admits a classical description of the local-hidden-state (LHS) model, namely,
\begin{equation}
\sigma_{a|x}=\sum_{\lambda}p(\lambda)p(a|x, \lambda)\rho'_{\lambda},~\forall~a,x
\label{eq_lhs_model}
\end{equation}
where $\{\rho'_{\lambda}\}$ is a set of preexisting quantum states, $p(\lambda)$ is a probability distribution, and $p(a|x,\lambda)$ denotes the postprocessing of Alice under the hidden variable $\lambda$. For a given assemblage, deciding whether it admits an LHS is a semidefinite program (SDP) \cite{steering_rpp_2017}. Whenever feasible solutions of Eq.~(\ref{eq_lhs_model}) can be found, the given assemblage admits an LHS and, by definition, is unsteerable; otherwise, steerable. The detailed constructions of the SDP can be found in Ref.~\cite{steering_rpp_2017}. Although an assemblage is generated by a quantum state, we stress that, it is the assemblage that conceived to be the intrinsic resource in quantum steering, rather than a state~\cite{Gallego2015PRX}. Throughout this work, we only consider the steerability from Alice to Bob, unless stated otherwise.

In contrast to the efficient verification of the steerability of an assemblage with an SDP, it becomes more difficult to verify the steerability of a bipartite quantum state with a given number of measurement settings $\#x=n$. To this end, one has to test all possible incompatible measurements with $\#x=n$. More specifically, to determine the least number of observables generating steerable assemblages, Alice must begin with iteratively testing two random observables until she successfully steers Bob; otherwise, she will measure one more observable if she concludes that she has no way to steer Bob with two observables, and so on. If the state requires at least $n$ observables from Alice to exhibit steerability on Bob, a quantum state is defined to be $n$-measurement steerable ($n$-MS), namely
\begin{equation}
n\textrm{-}\mathrm{MS}=\left\{\rho^\mathrm{AB}|\exists n\in\mathbb{N}\ni\left(\forall~\{\sigma_{a|x}\}_{a,x}\notin\textrm{LHS}\Rightarrow \#x\geq n\right)\right\}.
\end{equation}
The classification of states in terms of $n$-MS naturally forms a hierarchical structure that we define as the hierarchy of steering measurement settings. We remark that a similar issue has been raised in Bell nonlocality as well~\cite{brunner_bell_nonlocal_rmp_2014}. Even for the simplest case (qubit-pair Werner state), the exact local and nonlocal boundary is still vague \cite{Hirsch2017quantum}.

There are several feasible criteria in the literature discriminating the (un)steerability of a qubit-pair $\rho^{\rm{AB}}$ state. The BHQB criterion \cite{bowles_bhqb_ste_cri_pra_2016} is a sufficient condition for certifying the qubit-pair unsteerability; however, it is not strong enough to detect all unsteerable states. Additionally, two criteria of 2- and 3-MS are derived based on the CJWR steering inequality~\cite{PhysRevA.80.032112,PhysRevA.93.020103}. These 2- and 3-MS criteria are hardly generalized to the cases for $n\geq4$; moreover, they have a limited performance in detecting all 2- and 3-MS states. This insufficiency has been experimentally verified recently \cite{Adam2021PRA}. A brief review of these criteria is presented in Supplementary Note 1.

In light of the insufficiencies of aforementioned theoretical works~\cite{PhysRevA.80.032112,bowles_bhqb_ste_cri_pra_2016,PhysRevA.93.020103}, a computable approach to determine the hierarchy is desirable. We propose to leverage the compelling approach of supervised deep learning algorithms, which are clever at extracting hidden patterns and relationships within a huge amount of data.

\subsection*{Training data and SDP iteration}\label{sec_training data_sdp_iteration}

\begin{figure}[!ht]
\centering
\includegraphics[width=\textwidth]{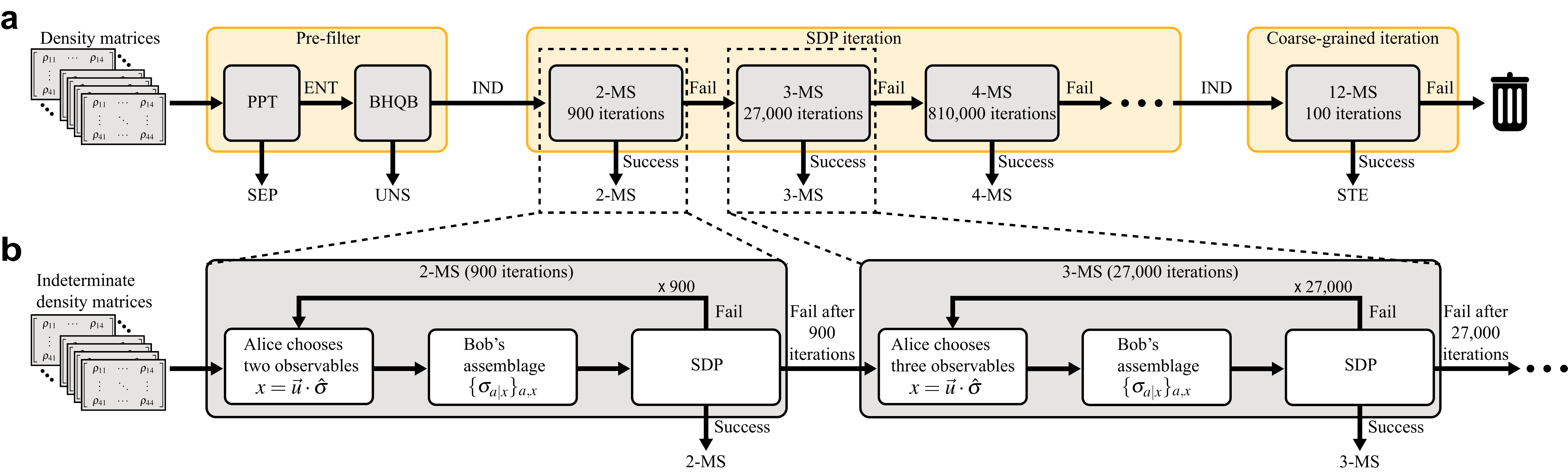}
\caption{\textbf{Computational protocol generating the labelled training data}. \textbf{a} To determine a reliable label $l$ as the ground truth of the hierarchy, each state is fed into a computational protocol, which consists of three stages of the pre-filter, the semidefinite program (SDP) iteration, and the coarse-grained iteration. The pre-filter consists of the positive partial transpose (PPT) and the Bowles-Hirsch-Quintino-Brunner (BHQB) criteria, which are used to efficiently capture the separable (SEP) and some unsteerable (UNS) states from the randomly generated qubit-pair density matrices fed into the protocol. After passing through the pre-filter, the states are entangled (ENT) with indeterminate (IND) steerability. In the final stage of the coarse-grained iteration, 12 random observables over 100 iterations has been worked out for sparse steerable (STE) states. \textbf{b} In the stage of the SDP iteration, the hierarchy can be determined. Each indeterminate input state will be iteratively tested by the SDP, under Alices's $n$ random observables $x = \vec{u}\cdot\hat{\sigma}$. Once the steerability is detected by the SDP for a certain set of $x$'s, then the state is $n$-measurement steerable ($n$-MS); otherwise, it will be repeatedly tested for a sufficiently long iteration and passed to the next level. The iteration times for 2-, 3-, and 4-MS states are 900, 27,000 and 810 thousands, respectively.}
\label{fig_sdp_iteration}
\end{figure}

In training a supervised deep learning model, it is necessary to generate a sufficient amount of labelled training data. Each training datum $(f,l)$ consists of the feature $f=(f_1,\ldots,f_k)\in\mathbb{R}^k$ of length $k$ encoding the object to be recognized and a label $l\in\mathbb{R}$ indicating the ground truth (GT) associated with the feature $f$. The goal of a deep learning model is to extract the relationship between $f$ and $l$ among the training data set $\{(f,l)\}$ and predict an $l^\prime$ for an $f^\prime$ never seen to the model. However, the theoretical insufficiencies discussed above prevent us from generating labelled training data efficiently, rendering the problem GT-deficient.

To overcome the GT-deficiency, we construct a computational protocol (Fig.~\ref{fig_sdp_iteration}\textbf{a}) to generate the necessary labelled training data. The first stage of the protocol is a pre-filter consisting of the positive partial transpose (PPT)~\cite{peres_ppt_prl_1996,pawel_ppt_pla_1997} and the BHQB criteria, which are used to efficiently capture the separable (SEP) and some unsteerable (UNS) states from the randomly generated qubit-pair density matrices fed into the protocol. After passing through the pre-filter, the states are entangled with indeterminate (IND) steerability.

The next stage is a fine-grained iterative SDP test determining the hierarchy. The detailed procedure is exemplified in Fig.~\ref{fig_sdp_iteration}\textbf{b} for the cases of 2- and 3-MS.
At each level of the hierarchy, the IND states will be iteratively tested by the SDP under $n$ randomly chosen observables $x=\vec{u}\cdot\hat{\sigma}$ by Alice, where $\vec{u}$ is a three-demensional unit vector. Then the states will be designated a label $l=n$-MS once the protocol successfully detects the steerability of the assemblage; otherwise, it will terminate after sufficiently many failures and pass the IND states to the next level. Note that the iterative test will be repeated for sufficiently many times to implement the optimization on Alice. Further details and statistical analyses are presented in Methods and Supplementary Note 2. Due to the limitation on our computational resource, here we have worked out a maximal $n=4$. To further promote the capability of the deep learning model, the final stage is the coarse-grained iteration, where a 12-MS over 100 iterations has been worked out (Fig.~\ref{fig_sdp_iteration}\textbf{a}). For those fail in this stage, we can not assign a reliable label. They are dropped without being appended to the training data.

\subsection*{Feature engineering}

Before feeding into the deep learning model, it is vital to properly encode the object to be recognized into a feature $f=(f_1,\ldots,f_k)\in\mathbb{R}^k$ of length $k$.
Since feature engineering reflects how the model recognizes the object, it has a significant influence on the behavior of the model.
Generically, the reduction of the feature length will facilitate promoting efficiency during the training;
nevertheless, this allows the model to acquire only limited information on the object, leading to a reduced prediction accuracy.

Here we consider five types of feature of different lengths, which is reduced based on the physical insights into quantum steering.
In contrast to the above generic intuition, we will see that the physics-driven reduction of the feature length is helpful in improving prediction accuracy.
This helps us to extract informative parameters relevant to quantum steering and to discover the physics behind the data.
Further details are presented in Methods.

Recall that, a qubit-pair state $\rho^\mathrm{AB}=\frac{1}{4}\sum_{ij}\Theta_{ij}\sigma_i\otimes\sigma_j$ can be fully described by $\Theta_{ij}=\Tr(\rho^\mathrm{AB}\sigma_i\otimes\sigma_j)$, where $\sigma_0=I_d$ and $\sigma_i$ denotes the three Pauli operators. Consequently, neglecting the triviality $\Theta_{00}=1$, $\rho^{\rm{AB}}$ is encoded into a feature of length $k=15$ consisting of $\Theta_{ij}$, denoted as General-15. Moreover, the state can be transformed to $\tilde{\rho}^\mathrm{AB}=[I_d\otimes(\rho^\mathrm{B})^{-1/2}]\rho^{\rm{AB}}[I_d\otimes(\rho^\mathrm{B})^{-1/2}]/2$
through the one-way stochastic local operations and classical communication (1W-SLOCC) on Bob, which preserves the steerability from Alice to Bob~\cite{bowles_bhqb_ste_cri_pra_2016,PhysRevA.90.024302,PhysRevA.92.032107,Ku2022NC}.
The denominator $2$ denotes the success probability of the transformation. It is critical to note that, since Bob's local state of $\tilde{\rho}^\mathrm{AB}$ is maximally mixed, we only need a feature of length $k=12$ to encode a qubit-pair state, denoted as SLOCC-12.

Additionally, a particularly inspiring technique studying quantum correlations is the quantum steering ellipsoid \cite{jevtic_steering_ellipsoid_prl_2014}.
Operationally, it is defined as the set of all attainable local states generated by measurements from the other side. For instance, if Alice performs all possible measurements, Bob's (normalized) local states in a Bloch sphere form an ellipsoid. In other words, Bob's steering ellipsoid contains all possible quantum states that Alice can steer to by her local measurements. This is reminiscent of the operational definition of the Alice-to-Bob steerability discussed in the previous section of Hierarchy of steering measurement settings.
The mathematical description of Bob's ellipsoid requires a $3\times3$ symmetric matrix $Q_\mathrm{B}$ and its center $\vec{c}_\mathrm{B}\in\mathbb{R}^3$, giving rise to a feature of length $k=9$, which we designate as ELLB-9; similarly, ELLA-9 for Alice. Furthermore, it has been pointed out \cite{jevtic_steering_ellipsoid_prl_2014} that both $\rho^\mathrm{AB}$ and $\tilde{\rho}^\mathrm{AB}$ have the same Alice's ellipsoid, underpinning the adequacy of ELLA-9 in the characterization of quantum steering.

Generically, the ellipsoids are obliquely aligned. We can further rotate the Bloch sphere such that the three semiaxes of Alices's ellipsoid are regularly aligned with the computational bases by applying an appropriate local unitary transformation on both sides. Note that the local unitary transformations preserve the steerability of a given quantum state. This corresponds to the elimination of the information on the orientation of the ellipsoid by diagonalizing $Q_\mathrm{A}$, leading to the most compact feature of length $k=6$, denoted as LUTA-6. According to our findings below, here we merely consider Alice's regularly aligned ellipsoid.

\subsection*{Supervised deep learning model}

\begin{figure}[!ht]
\centering
\includegraphics[width=0.6\textwidth]{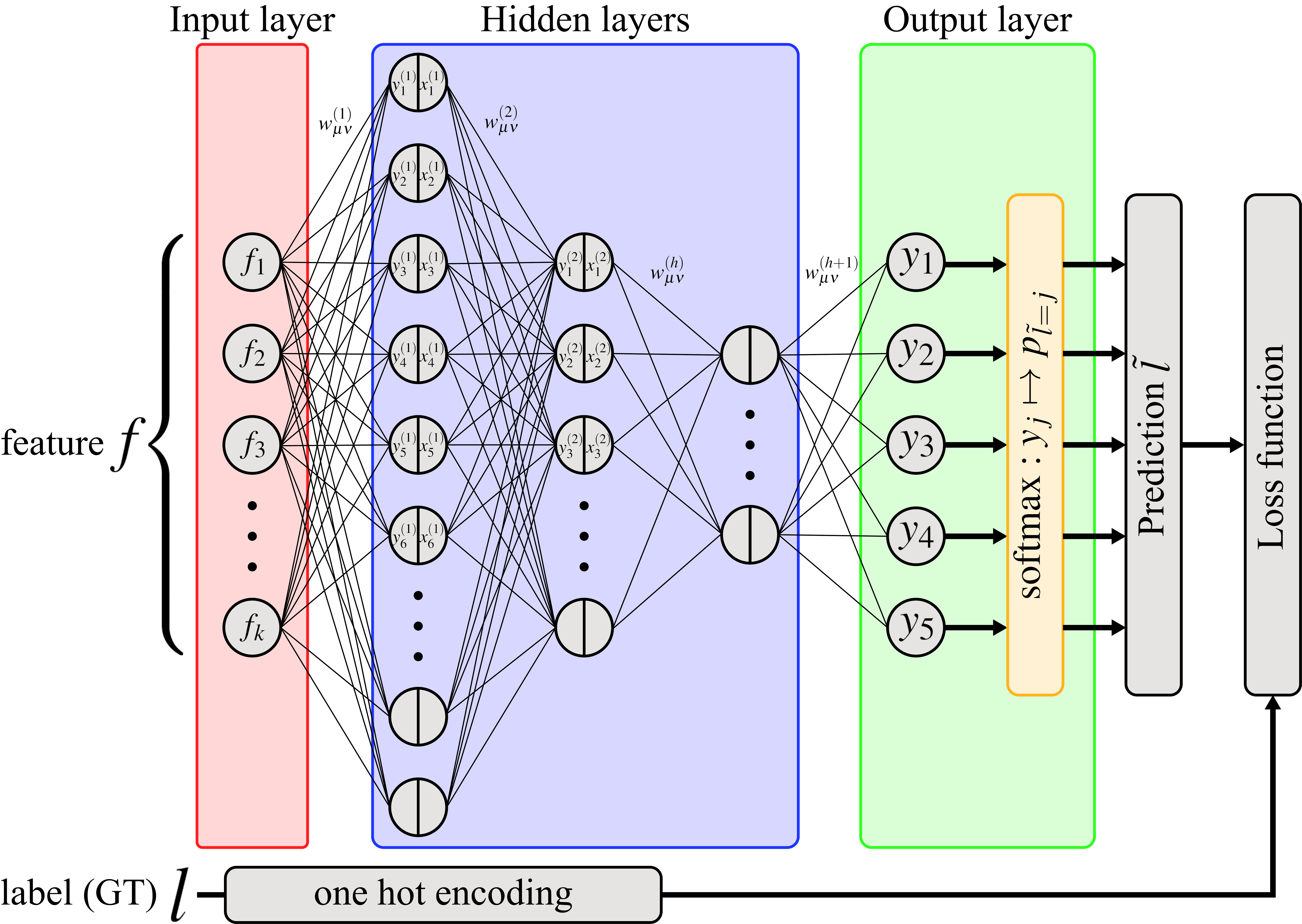}
\caption{\textbf{The structure of an artificial neural network (ANN)}. An ANN consists of an input layer, multiple hidden layers, and an output layer.
The weights $\{w^{(m)}_{\mu\nu}\}$ connecting the neurons between layers are optimized with the conventional backpropagation algorithm aiming at minimizing the loss function.
The prediction $\tilde{l}$ is determined according to the probability given by the softmax function via the five outputs from the output layer. Then $\tilde{l}$ will be compared with the ground truth (GT) $l$ to estimate the loss function, which in turn determines how the weights $\{w^{(m)}_{\mu\nu}\}$ are updated.}
\label{fig_ann_structure}
\end{figure}

The artificial neural network (ANN) is one of the most widely used learning models. We have trained five ANN models, each of which is optimized for one type of features accordingly.
As schematically shown in Fig.~\ref{fig_ann_structure}, an ANN consists of an input layer, multiple hidden layers, and an output layer.
The number of neurons in the input layer is the same as the feature length $k$. In our constructions, the detailed structure of the hidden layers will be optimized for each type of features.
As the ANN models are trained to forecast the five types of labels, the output layer is composed of five neurons followed by the softmax function outputting the probability of each label. The most probable prediction $\tilde{l}$ can be determined accordingly.
The weights $\{w^{(m)}_{\mu\nu}\}$ connecting the neurons between layers are optimized with the conventional backpropagation algorithm aiming at minimizing the loss function.
And the outputs of the neurons in the hidden layers are determined by an activation function $\mathrm{ReLU}:y^{(m)}_\mu\mapsto x^{(m)}_\mu$.
Further details on the structure and implementation of the ANN models are presented in Supplementary Note 4.

\subsection*{Verification of the accuracy of the models}

\begin{figure}[!ht]
\centering
\includegraphics[width=\textwidth]{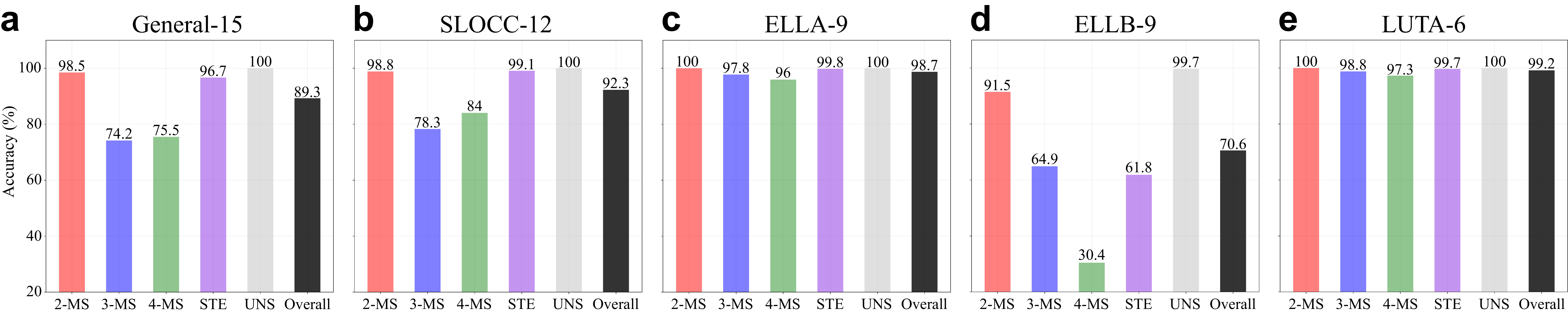}
\caption{\textbf{The accuracy of the well-trained models on the testing data}. The accuracy for each label, i.e., $n$-measurement steerable ($n$-MS), steerable (STE) with $n\geqslant5$, and unsteerable (UNS), on the 10,300 testing data along with the different features, including General-15, stochastic local operation with classical communication (SLOCC-12), Alice's steering ellipsoid (ELLA-9), Bob's steering ellipsoid (ELLB-9), and Alice's regularly aligned ellipsoid (LUTA-6), is shown in panels \textbf{a} to \textbf{e}. Crucially, except for the ELLB-9 model, the overall performance of the model is significantly improved, indicating an even compact, and informative, encoding scheme. Particularly, the LUTA-6 model is capable of predicting the hierarchy of steering measurement settings at a very high overall accuracy of 99.2 $\%$. On the other hand, the ELLB-9 model shows an unsatisfactory accuracy even though this feature encodes all of Bob's states that can be steered to from Alice. From
these responses of the models to different features, we can pinpoint that the adequate characterization of the steerability from Alice to Bob should be given by the Alice's regularly aligned ellipsoid.}
\label{fig_dl_pred_accuracy}
\end{figure}

Now we proceed to verify the accuracy of the well-trained models. The testing data of size 10,300 are also output from the protocol other than those in the training data set.
The details on the grouping of the generated data into training and test sets are presented in Supplementary Note 3.
It will become clear that, from the responses of the models to the physics-driven features, we can figure out the most informative parameters relevant to the characterization of quantum steering.
More specifically, a counterintuitive conclusion can be drawn that the steerability from Alice to Bob is dominantly characterized by Alice's regularly aligned ellipsoid.

Figure~\ref{fig_dl_pred_accuracy} shows the accuracy of the models trained with each feature. It can be seen that, except for the ELLB-9 model shown in Fig.~\ref{fig_dl_pred_accuracy}\textbf{d}, the overall accuracy gets improved significantly along with the reduced features. Particularly, the LUTA-6 model shown in Fig.~\ref{fig_dl_pred_accuracy}\textbf{e} almost precisely reproduces all the labels of the testing data. This, on the one hand, implies that the six parameters of the LUTA-6 feature dominantly characterize the steerability from Alice to Bob. On the other hand, even if the General-15 feature does contain complete information on the input states $\rho^\mathrm{AB}$, the redundancy in the feature irrelevant to the steerabiltiy expands the model space. This hinders the trainability of the General-15 model, giving rise to a limited accuracy shown in Fig.~\ref{fig_dl_pred_accuracy}\textbf{a}.

We then take a further closer look into the procedure of feature engineering to reveal deeper physical insights into the behavior of the models. The feature reduction begins with the 1W-SLOCC transformation $\rho^\mathrm{AB}\mapsto\tilde{\rho}^\mathrm{AB}=[I_d\otimes(\rho^\mathrm{B})^{-1/2}]\rho^{\mathrm{AB}}[I_d\otimes(\rho^\mathrm{B})^{-1/2}]/2$, which is shown to preserve the steerability from Alice to Bob~\cite{bowles_bhqb_ste_cri_pra_2016,PhysRevA.92.032107,Nery2020,Ku2022NC}, leading to a more compact SLOCC-12 model with improved accuracy shown in Fig.~\ref{fig_dl_pred_accuracy}\textbf{b}.

The aforementioned technique of quantum steering ellipsoid \cite{jevtic_steering_ellipsoid_prl_2014} characterizes the entanglement in a geometrical manner; whereas, it remains vague how can this approach characterize quantum steering. In Fig.~\ref{fig_dl_pred_accuracy}\textbf{c} (\textbf{d}), we show the results of the ELLA(B)-9 models, encoded based on the ellipsoids of Alice (Bob), respectively. On the one hand, the ELLA-9 model exhibits further improved accuracy than the SLOCC-12, indicating an even further compact encoding scheme, and less redundancy.

On the other hand, the failure of ELLB-9 is seemingly counterintuitive. Since Bob's ellipsoid is defined in the similar way of Alice-to-Bob steerability, i.e., the set of all states $\rho^{\rm{B}}_{a|x}$ that can be steered to from Alice, an intuitive conjecture arises naturally that Bob's ellipsoid should be adequate to characterize the steerability from Alice to Bob. There are also attempts in the literature to connect both concepts, e.g., the semiaxes of Bob's steering ellipsoid can be used to determine Alice's optimal measurements~\cite{McCloskey2017PRA,Ku2018PRA,song_njp_2023}. Although the accuracy of the ELLB-9 model is unsatisfactory,
it provides information about hidden steerability, which will be discussed later.

The physics behind the behaviors of the ELLA-9 and the ELLB-9 models can be both understood from the critical properties of the 1W-SLOCC. Since any 1W-SLOCCs on Bob can neither alter Alice's steering ellipsoid~\cite{jevtic_steering_ellipsoid_prl_2014}, nor the detection of steerable state from Alice to Bob~\cite{Nery2020,Ku2022NC}.
More generically, any 1W-SLOCC transformations \begin{equation}
\rho^\mathrm{AB}\mapsto\hat{\rho}^\mathrm{AB}=\frac{(I_d\otimes K)\rho^\mathrm{AB}(I_d\otimes K)^\dagger}{\mathrm{Tr}[(I_d\otimes K)\rho^\mathrm{AB}(I_d\otimes K)^\dagger]}
\end{equation}
do not alter neither the Alice-to-Bob steerability nor Alice's steering ellipsoid~\cite{jevtic_steering_ellipsoid_prl_2014,Ku2022NC,Hsieh2023arXiv}. Here, Kraus operator $K$ satisfying $K^{\dagger}K\leq I_d$ forms a valid quantum operation, and the denominator denotes the renormalization factor (or success probability of the transformation). Consequently, the ELLA-9 feature should contain all the necessary information to certify the steerability. We therefore put particular emphasize on the conclusion that, it should be Alice's steering ellipsoid, instead of Bob's ellipsoid, that adequate to characterize the steerability from Alice to Bob, underpinned by the further improved accuracy of the ELLA-9 model as well. In contrast, the 1W-SLOCCs on Alice do change the steerability from Alice to Bob. However, the ELLB-9 feature of a state is invariant under the 1W-SLOCCs on Alice. Namely, two states related to each other via a 1W-SLOCC on Alice should share the same ELLB-9 encoding with different steerability, rendering the label $l$ given by the protocol unreliable. This severely suppresses the accuracy of the ELLB-9 model. We will also demonstrate the effects of the 1W-SLOCCs on Alice in detecting the hidden steerability.

Finally, the LUTA-6 feature is inherited from Alice's ellipsoid by rotating the Bloch sphere such that it is regularly aligned, leading to the most compact feature and a highly precise prediction shown in Fig.~\ref{fig_dl_pred_accuracy}\textbf{e}. This highlights that the orientation of Alice's ellipsoid is irrelevant in the characterization of quantum steering. Specifically, from the above feature engineering, the steerability of a state
$\rho^\mathrm{AB}=\frac{1}{4}\sum_{ij}\Theta_{ij}\sigma_i\otimes\sigma_j$ is invariant under the cascased transformations
\begin{equation}
\rho^\mathrm{AB}\mapsto\tilde{\rho}^\mathrm{AB}\mapsto\rho^{\prime\mathrm{AB}}=(U^\mathrm{A}\otimes U^\mathrm{B})\tilde{\rho}^\mathrm{AB}(U^\mathrm{A}\otimes U^\mathrm{B})^\dagger,
\end{equation}
where $U^\mathrm{A}\otimes U^\mathrm{B}$ is an appropriate local unitary operator diagonalizing $Q_\mathrm{A}$.

From the above physics-based deduction and the high accuracy of the well-trained model, the following corollary can be drawn:
\begin{coro}
Alice's regularly aligned ellipsoid determines the most compact encoding of Alice-to-Bob steerability for qubit-pair states.
\end{coro}
It is also worthwhile to emphasize that this reasoning is independent of the hierarchy of steering measurement settings. Furthermore, since the effect of 1W-SLOCC operation on steerability is independent of the system dimension, it is also helpful in decreasing the necessary parameters characterizing the quantum steering in higher-dimensional cases. Note that, although the definition of a higher-dimensional steering ellipsoid is vague, decreasing the encoding parameters of the steerability by 1W-SLOCC remains valid in higher-dimensional cases.

This also explains the insufficiency of the CJWR-based criteria, which are used to detect the steerability of a qubit-pair state with 2 and 3 measurement settings~\cite{PhysRevA.93.020103}.
The CJWR-based criteria only take the eigenvalues of $Q_\mathrm{A}$ into account and neglect $\vec{c}_\mathrm{A}$, which we have shown to be critical as well in characterizing the steerability. In this sense, the CJWR-based criteria are considered to be symmetric. Further detailed behaviors of the models expressed in terms of the confusion matrices are shown in Supplementary Note 5.

\subsection*{Hierarchy predicted by the models and hidden steerability}

\begin{figure}[!ht]
\centering
\includegraphics[width=\textwidth]{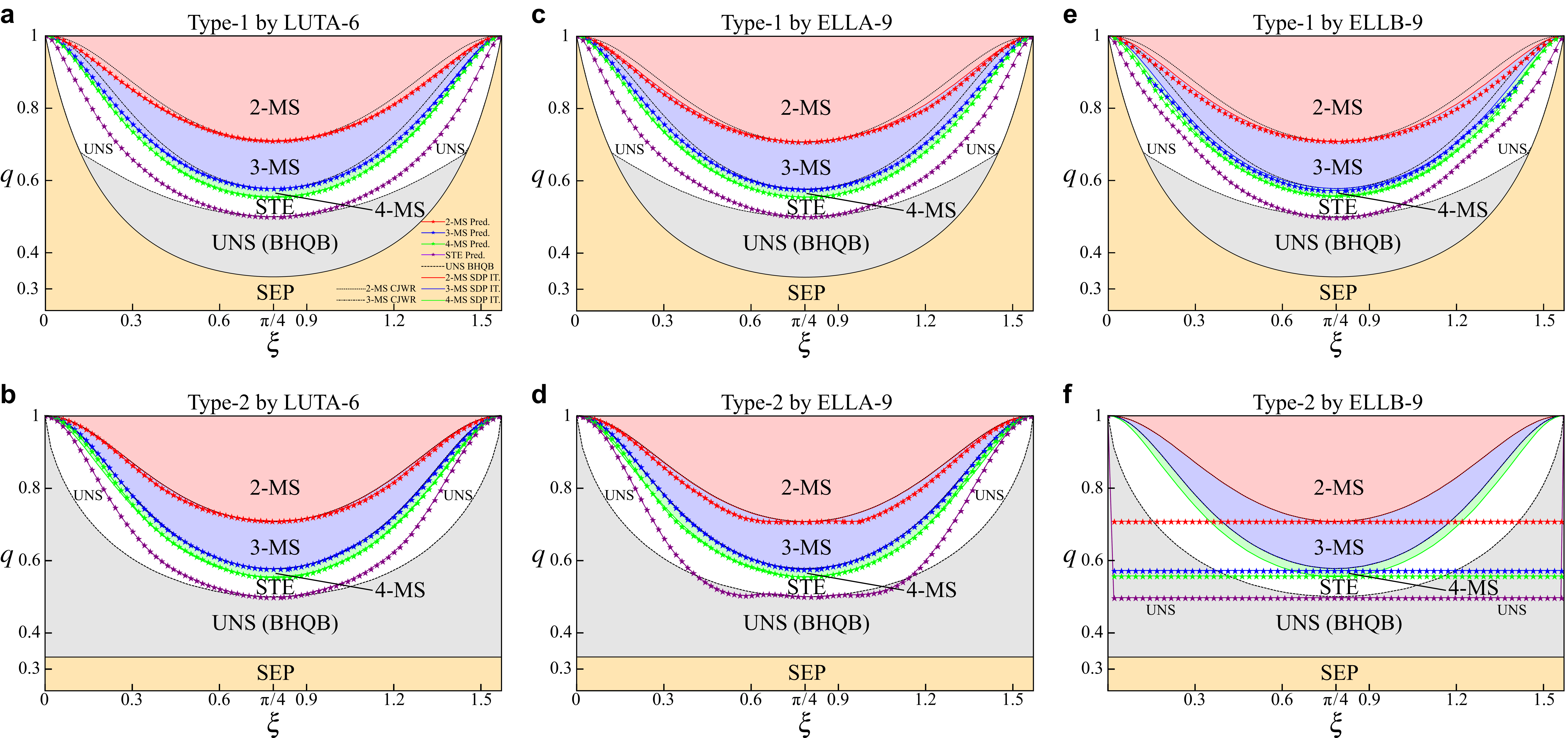}
\caption{\textbf{Prediction of the hierarchy of both types of states}. The predictions of the hierarchy (starry curves) for \textbf{a} type-1 $\rho_1(q,\xi)$ and \textbf{b} type-2 states $\rho_2(q,\xi)$ by the model trained with the feature of Alice's regularly aligned ellipsoid (LUTA-6) are in good agreement with the ground truth given by the protocol (solid curves). Due to its high accuracy, the LUTA-6 model is capable of discovering several physical insights. More specifically, in the case of type-1 states, the predictions deviate from the Cavalcanti-Jones-Wiseman-Reid (CJWR)-based criteria (dotted and dot-dashed curves), indicating the insufficiency of the existing theoretical approaches in the literature. Furthermore, for both types of states, the LUTA-6 model can predict the border of the green regions of 4-measurement steerable (MS) in high precision, which is hard to detect with existing theoretical approaches; meanwhile, we also show the predicted boundaries (purple starry curves) between steerable (STE) with $n\geqslant5$ and unsteerable (UNS), sandwiched between the 4-MS and the UNS detected by the Bowles-Hirsch-Quintino-Brunner (BHQB) criterion. In contrast, the predictions of the hierarchy (starry curves) by the model trained with the feature of Alice's ellipsoid (ELLA-9) (for \textbf{c} type-1 and \textbf{d} type-2 states) and the model trained with the feature of Bob's ellipsoid (ELLB-9) models (for \textbf{e} type-1 and \textbf{f} type-2 states) are worse, compared to the LUTA-6 counterpart. Although the overall profile of the predicted hierarchy for type-1 states remain in line with the ground truth, the ELLB-9 model appears to completely fail to predict the hierarchy of type-2 states. Whereas, we stress that, the predicted horizontal lines in turn reveal the underlying physics of hidden steerability. The orange region in each panel stands for separable (SEP) states which can be efficiently captured by the positive partial transpose criterion.}
\label{fig_dl_pred_states}
\end{figure}

To further showcase the merit of the well-trained models, meanwhile to facilitate the visualization of the predictions, we apply these models to predict the hierarchies of two different types of quantum states generalized from the standard Werner state; namely,
\begin{equation}
\begin{aligned}
\rho_1(q,\xi)=&q\ket{\Psi_\xi}\bra{\Psi_\xi}+(1-q)\frac{I_d}{2}\otimes\frac{I_d}{2}\\
\rho_2(q,\xi)=&q\ket{\Psi_\xi}\bra{\Psi_\xi}+(1-q)\rho^{\rm{A}}\otimes\frac{I_d}{2}.
\label{equation11}
\end{aligned}
\end{equation}
where $\ket{\Psi_\xi}=\cos\xi\ket{00}+\sin\xi\ket{11}$, $0\leq\xi\leq\pi/2$, $q\in[0,1]$, and $\rho^{\mathrm{A}}=\mathrm{Tr}_{\mathrm{B}}\ket{\Psi_\xi}\bra{\Psi_\xi}$. Here we use the subscribes in $\rho_1$ and $\rho_2$ to denote the two different types of noise adding to the pure entangled state $\ket{\Psi_\xi}$.

In view of the accuracy presented in Fig.~\ref{fig_dl_pred_accuracy}, we show the hierarchies predicted by the LUTA-6, the ELLA-9, and the ELLB-9 models (starry curves) in Fig.~\ref{fig_dl_pred_states}
for comparison. Additionally, it is also insightful to compare the model predictions with the results output from the protocol (colored regions) explained in Fig.~\ref{fig_sdp_iteration}, which serve as
the GT to evaluate the model performance on these two types of states. To provide a quantitative evaluation of the model performance, we also estimate the mean absolute displacement (MAD)
of the border predicted by the model from that given by the protocol according to
\begin{equation}
\mathrm{MAD}=\frac{\sum_{\xi=0}^{\pi/2} |q_\xi^{(\mathrm{Prediction})}-q_\xi^{(\mathrm{GT})}|}{\mathrm{number~of~pixels~on~a~border}}.
\end{equation}
The numerical results are shown in Table~\ref{tab_mad}. Further predictions by the General-15 and the SLOCC-12 models are presented in Supplementary Note 6.

Figures~\ref{fig_dl_pred_states}\textbf{a} and \textbf{b} show the predictions by the LUTA-6 model for the type-1 and type-2 states, respectively. As expected, the predicted boundaries between 2- and
3-MS, as well as between 3- and 4-MS, are in good agreement with the fine-grained hierarchies given by the protocol, leading to the very small MAD values of the LUTA-6 model for both types
of states in Table~\ref{tab_mad}. Additionally, the deviations from those given by the CJWR-based criteria (dotted and dot-dashed curves) can also be clearly observed in
Fig.~\ref{fig_dl_pred_states}\textbf{a}. This discrepancy is in line with the recent experimental report~\cite{Adam2021PRA}, wherein merely 2 or 3 Pauli measurements have been tested $\rho_1(q,\xi)$.

\begin{table}
\caption{\textbf{Quantitative evaluation of the model performance}. We propose to quantitatively evaluate the model performance in terms of the mean absolute displacement (MAD) of the border of $n$-measurement steerable ($n$-MS) states predicted by the
model from that given by the protocol. Here we show the numerical results of the LUTA-6, the ELLA-9, and the ELLB-9 models on the two types of states.}
\begin{tabular}{c|ccc}
\hline\hline
Type-1& LUTA-6              & ELLA-9              & ELLB-9 \\
\hline
2-MS  & $2.13\times10^{-3}$ & $4.47\times10^{-3}$ & $7.67\times10^{-3}$ \\
3-MS  & $0.44\times10^{-3}$ & $0.71\times10^{-3}$ & $7.82\times10^{-3}$ \\
4-MS  & $1.24\times10^{-3}$ & $1.11\times10^{-3}$ & $4.59\times10^{-3}$ \\
\hline\hline
Type-2& LUTA-6              & ELLA-9              & ELLB-9 \\
\hline
2-MS  & $4.64\times10^{-3}$ & $11.06\times10^{-3}$ & $125\times10^{-3}$ \\
3-MS  & $4.8 \times10^{-3}$ & $1.24\times10^{-3}$ & $172\times10^{-3}$ \\
4-MS  & $5.94\times10^{-3}$ & $3.72\times10^{-3}$ & $163\times10^{-3}$ \\
\hline\hline
\end{tabular}
\label{tab_mad}
\end{table}

Furthermore, the green regions of 4-MS states, which are hardly detected by the existing theoretical approaches in the literature, have also been predicted by the LUTA-6 model in high precision.
According to our results, the range of the 4-MS standard Werner state $\rho_\mathrm{W}(q)=\rho_1(q,\pi/4)$ is $1/\sqrt{3}>q>0.555$. Inferring from the hierarchical structure shown in Fig.~\ref{fig_dl_pred_states}, it is reasonable to conjecture that each $n$-MS band would be narrower with increasing $n$. In addition, we also show the potential STE-UNS boundaries (purple starry curves) suggested by the models, lying within the blank region of IND states sandwiched between the 4-MS and the UNS detected by the BHQB criterion. This also reflects the insufficiency of the BHQB criterion.

The results predicted by the ELLA-9 model for the type-1 and type-2 states are shonw in Figs.~\ref{fig_dl_pred_states}\textbf{c} and \textbf{d}, respectively.
As expected, the overall performance of the ELLA-9 model is slightly lower than that of the LUTA-6 model. However, the predicted boundaries for the type-1 states still fit well into the fine-grained
hierarchies given by the protocol. For the case of type-2 states, except for the 2-MS border, the other two predictions are even numerically better than that of LUAT-6 model.

Compared to the previous two models, the performance of the ELLB-9 model (Figs.~\ref{fig_dl_pred_states}\textbf{e} and \textbf{f}) is worse, as numerically shown in Table~\ref{tab_mad}.
It is interesting to note that, actually, type-1 states result in the same ellipsoids for Alice and Bob due to the high symmetry of the states;
therefore, the overall profile of the predicted hierarchy for type-1 states remains in line with the GT. On the other hand, the ELLB-9 model appears to completely fail to predict the hierarchy of type-2 states. Every state is identified as the Werner state $\rho_\mathrm{W}(q)=\rho_2(q,\pi/4)$, giving rise to the horizontal borders.

To understand the seemingly erroneous horizontal borders given by the ELLB-9 model for the type-2 states (Fig.~\ref{fig_dl_pred_states}\textbf{f}), we notice that each type-2 state $\rho_2(q,\xi)$ will be transformed to the Werner state by an 1W-SLOCC on Alice, i.e., $\rho_\mathrm{W}(q)=[(\rho^{\rm{A}})^{-1/2}\otimes I_d] \rho_2(q,\xi)[(\rho^{\rm{A}})^{-1/2}\otimes I_d]/2$, $\forall~\xi\in[0,\pi/2]$.
In other words, for a given $q$, the states $\rho_2(q,\xi)$ lying on the same horizontal line share the same ELLB-9 feature encoding with $\rho_\mathrm{W}(q)$, and, consequently, are identified as the Werner state $\rho_\mathrm{W}(q)$ by the ELLB-9 model. However, $\rho_2(q,\xi)$ and $\rho_\mathrm{W}(q)$ do exhibit different hierarchies of steering measurement settings. This, on the one hand, results in the seemingly erroneous prediction shown in Fig.~\ref{fig_dl_pred_states}\textbf{f}. On the other hand, the inconsistency of the predictions by the LUTA-6 and ELLB-9 models for type-2 states (Figs.~\ref{fig_dl_pred_states}\textbf{b} and \textbf{f}) is reminiscent of the famous concept of hidden steerability, where an unsteerable state can become steerable w.r.t. an appropriate 1W-SLOCC on Alice~\cite{PhysRevA.92.032107,Quintino2016PRA,Hsieh2016PRA,Pramanik2019}. In particular, in the previous studies, the SLOCCs on both Alice and Bob were considered to activate hidden steering. The prediction from deep learning model suggests that it is sufficient to consider 1W-SLOCC on Alice to activate hidden steering~\cite{Ku2023arXiv,Hsieh2023arXiv}.

\section*{Discussion}\label{sec_discussion}

Although the assemblage is known to be the intrinsic resource in quantum steering, its optimal preparation from a bipartite state remains unclear. We attempt to shed light on the avenue toward this issue. In this work, we tackle the cumbersome optimization over all possible incompatible measurements with two approaches. Each of which provides deeper insights into the characterization of quantum steering, leading to the hierarchy of steering measurement settings.

We first develop a computational protocol implementing the optimization by sufficiently long iterations under $n$ random observables. We concretely exemplify the protocol with two types of generalized Werner state, and clearly showcase its merit in overcoming the insufficiencies of the theoretical criteria in the literature, including the fine-grained 2- and 3-MS borders and a 4-MS region inaccessible to the existing criteria.

We also manage to infer the fine-grained hierarchy of steering measurement settings by leveraging the power of the supervised deep learning algorithm. According to the responses of the well-trained models to the physics-driven features, we can acquire deeper physical insights seemingly contradictory to existing operational definitions. Our results suggest that, Alice’s regularly aligned steering ellipsoid is the most compact characterize of Alice-to-Bob steerability for qubit-pair states. Additionally, we also showcase the versatility of our models in predicting the hierarchy and the exploration of the hidden quantum steering.

Our results naturally open several new directions. For example, an explicit characterization of steerability in terms of Alice’s ellipsoid would be impactful, particularly in constructing stronger steering criteria in qubit-pair scenarios. For a given qubit-pair state, our protocol can be helpful in finding the optimal preparation of steerable assembles. To explore a compact characterization of two-way steerability in a similar way would be a heuristic attempt. Crucially, our results also underpin the possibility of the deep learning algorithms in mining new physics behind a huge amount of data, rather than a substitute of cumbersome computational task.

($\equiv\widehat{\mathsf{\Phi}}\omega\widehat{\mathsf{\Phi}}\equiv$) $\sim$ meow

\section*{Methods}\label{sec_method}

\subsection*{Data generation and SDP iteration}

The training data set $\{(f,l)\}$ is a collection of tuples, where $f\in\mathbb{R}^k$ is the feature encoding the object to be recognized and $l\in\mathbb{R}$ is the label indicating the ground truth associated to the feature $f$. Due to the GT-deficient nature of the problem to be resolved, we  construct a computational protocol to determine the correct label $l$ of each datum, i.e., the hierarchy of steering measurement settings of qubit-pair states.

By running the \texttt{RandomDensityMatrix} code~\cite{RandomDensityMatrix}, we can efficiently generate a large amount of random density matrices (Fig.~\ref{fig_sdp_iteration}\textbf{a}). Then we feed them into the protocol described below to determine the label $l$ of each state.

The first stage of the protocol is a pre-filter, which consists of two discriminators capable of efficiently capturing those states manifestly unsteerable. The PPT criterion can efficiently and exactly determine the separability ($l=\mathrm{SEP}$) of a qubit-pair state and pass the entangled state to the next BHQB criterion. Although the BHQB criterion is sufficient for unsteerability ($l=\mathrm{UNS}$), it can be efficiently implemented and enhance the ratio of steerable states among the output states from the pre-filter. After passing through the pre-filter, the states are entangled with indeterminate (IND) steerability.

The second stage is the SDP iteration used to determine the label of $n$-MS. The detailed procedure is schematically shown in Fig.~\ref{fig_sdp_iteration}\textbf{b}. The IND states from the pre-filter are sent to the first level of 2-MS, wherein Alice will randomly perform two observables. Then Bob obtains the corresponding assemblages and determines their steerability with SDP. Once the assemblages are steerable, then the corresponding states are designated a label $l=2$-MS. For those fail to demonstrate steerability, Alice will perform the measurements again with two new random observables. This SDP test will be repeated for at most 900 times at the level of 2-MS. The states fail to demonstrate 2-MS after 900 iterations will be sent to the next level of 3-MS. At each level of the hierarchy, the procedure is similar besides the number $n$ of Alice's observables and the times of iterative test. For the following level of 3-, 4-, and 5-MS, the iterative test will be repeated for 27,000, 810 thousands, and 24 millions times, respectively. Such high repetition is to ensure the optimization over all possible incompatible measurements on Alice's side. Further statistical analyses on the iteration times are presented in Supplementary Note 2.

It can be seen that, to ensure the optimization, the iteration times increases exponentially with the number $n$ of observables. The demand on the computing power rapidly exceeds our machines. Here we have worked out a maximal $n=4$. However, it is manifest that there should be steerable states with $n>4$, which can not be detected with the fine-grained SDP iteration terminating at 4-MS in the second stage. In the final stage of coarse-grained iteration (Fig.~\ref{fig_sdp_iteration}\textbf{a}), we manage to dig out more steerable states to the largest extent from the indeterminate states fail at level $n=4$. The procedure of this stage is also the same as those in the SDP iteration. But here, limited by our computational resource, we set $n=12$ and repeat for at most 100 times. For states successfully demonstrate steerability, we designate a label $l=\mathrm{STE}$. On the other hand, for those fail in this stage, we can not decide whether they are steerable with $n\geq13$ or unsteerable. Consequently, we can not designate a reliable label and drop them from the data set.

\subsection*{Extracting relevant parameters in the feature}

Here we present the construction of five types of feature encoding a qubit-pair state $\rho^\mathrm{AB}$, and explain how to extract informative parameters relevant to the characterization of quantum steering, leading to a compact feature.

As discussed in the main text, the density matrix of a qubit-pair state $\rho^\mathrm{AB}=\frac{1}{4}\sum_{ij}\Theta_{ij}\sigma_i\otimes\sigma_j$ can be expressed in terms of a $4\times4$ real matrix $\Theta$ with $\Theta_{ij}=\Tr(\rho^\mathrm{AB}\sigma_i\otimes\sigma_j)$, where $\sigma_0=I_d$ and $\sigma_i$ denotes the three Pauli operators. It can be seen that $\Theta$ has a block structure:
\begin{equation}
\Theta=\left[\begin{array}{cc}
1 & \vec{b}^{T} \\
\vec{a} & T \\\end{array}\right].
\end{equation}
Since any density matrices are always of unital trace, $\Theta_{00}=1$ is trivial and irrelevant to quantum steering. Therefore, we need 15 real parameters to fully describe $\rho^\mathrm{AB}$, denoted as General-15.

Consider the one-way stochastic local operations and classical communication (1W-SLOCC) on Bob via the mapping:
\begin{equation}
\rho^\mathrm{AB}\mapsto\tilde{\rho}^\mathrm{AB}=[I_d\otimes(\rho^\mathrm{B})^{-1/2}]\rho^{\rm{AB}}[I_d\otimes(\rho^\mathrm{B})^{-1/2}]/2.
\label{eq_1w-slocc}
\end{equation}
It is critical to note that Bob's local state of $\tilde{\rho}^\mathrm{AB}$ is maximally mixed, corresponding to the transformed matrix
\begin{equation}
\widetilde{\Theta}=\left[\begin{array}{cc}
1 & 0^T \\
\tilde{a} & \widetilde{T} \\\end{array}\right].
\label{eq_theta_slocc}
\end{equation}
More specifically, $\widetilde{T}$ is a $3\times3$ real matrix determined by $\widetilde{T}_{ij}=\Tr(\tilde{\rho}^\mathrm{AB}\sigma_i\otimes\sigma_j)$ for $i.j\in\{1,2,3\}$, and $\tilde{a}\in\mathbb{R}^3$.
Therefore, we have only 12 nontrivial parameters left, denoted as SLOCC-12. Moreover, it has been pointed out~\cite{bowles_bhqb_ste_cri_pra_2016,PhysRevA.90.024302,PhysRevA.92.032107,Ku2022NC} that $\tilde{\rho}^\mathrm{AB}$ and $\rho^\mathrm{AB}$ have the same steerability from Alice to Bob, provided that $\rho^\mathrm{B}=\Tr_{\rm{A}}\rho^{\rm{AB}}$ is a mixed state. Therefore, SLOCC-12 is capable of characterizing quantum steerability with lesser parameters.

Inspired by the technique of quantum steering ellipsoid \cite{jevtic_steering_ellipsoid_prl_2014}, we would like to ask whether quantum steering can also be characterized in a geometrical manner.
Suppose that Bob performs a measurement described by an operator $E=\frac{1}{2}\sum_{j=0}^{3}X_j\sigma_j$, then Alice will obtain her local state $\frac{1}{2}\Theta X$ with probability $p_E=\frac{1}{2}(1+\vec{b}\cdot\vec{X})$. According to the operational definition, Alice's ellipsoid is obtained by collecting the normalized local state $\frac{1}{2}\Theta X/p_E$ for all possible measurement operators $E$. Moreover, it has been pointed out~\cite{jevtic_steering_ellipsoid_prl_2014} that Alice’s ellipsoid is invariant under the 1W-SLOCC on Bob (\ref{eq_1w-slocc}). This provides a convenient way for expressing Alice’s ellipsoid for $\rho^\mathrm{AB}$ with the parameters in the transformed matrix $\widetilde{\Theta}$ in Eq.~(\ref{eq_theta_slocc}); namely,
\begin{equation}
Q_\mathrm{A}=\widetilde{T}\widetilde{T}^T.
\label{eq_ellipsoid_qa}
\end{equation}
Note that $Q_\mathrm{A}$ is a $3\times3$ symmetric matrix; therefore, along with the center of the ellipsoid $\vec{c}_\mathrm{A}=\tilde{a}$, we need merely at most 9 parameters in total to encode $\rho^\mathrm{AB}$, denoted as ELLA-9. Similar procedure can be done with respect to the 1W-SLOCC operator $(2\rho^\mathrm{A})^{-1/2}\otimes I_d$ on Alice, leading to the ELLB-9. This reduction is achieved by symmetrizing $\widetilde{T}$ by $Q_\mathrm{A}$ in Eq.~(\ref{eq_ellipsoid_qa}).

For a randomly generated $\rho^\mathrm{AB}$, Alice's ellipsoid may be obliquely aligned; namely, the off-diagonal elements of $\widetilde{T}$ may be non-zero. To construct the most compact feature, we can diagonalize $\widetilde{T}$ by applying an appropriate local unitary transformation on both sides as
\begin{equation}
\tilde{\rho}^\mathrm{AB}\mapsto\rho^{\prime\mathrm{AB}}=(U^\mathrm{A}\otimes U^\mathrm{B})\tilde{\rho}^\mathrm{AB}(U^\mathrm{A}\otimes U^\mathrm{B})^\dagger.
\end{equation}
Then we have
\begin{equation}
\Theta^\prime=\left[\begin{array}{cc}
1 & 0^T \\
\vec{a}^\prime & T^\prime \\\end{array}\right]
\end{equation}
with $T^\prime$ being a diagonal matrix, leading to the most compact feature with $k=6$, denoted as LUTA-6. In this construction, the information on the orientation of the ellipsoid is removed.

\section*{Data availability}

The data that support the findings of this study are available upon reasonable request from the corresponding authors.

\section*{Code availability}

The code that supports the findings of this study is available upon reasonable request from the corresponding authors.


\section*{Acknowledgments}

The authors acknowledge fruitful discussions with Macro T. Quantino, Yi-Te Huang, Shin-Liang Chen, Che-Ming Li, and Costantino Budroni.
This work is supported by the National Science and Technology Council, Taiwan, with Grants No. MOST 108-2112-M-006-020-MY2, MOST 109-2112-M-006-012, MOST 110-2112-M-006-012, and MOST 111-2112-M-006-015-MY3,
partially by the Higher Education Sprout Project, Ministry of Education to the Headquarters of University Advancement at NCKU,
and partially by the National Center for Theoretical Sciences, Taiwan.
H.-Y.K. is supported by the National Center for Theoretical Sciences and the National Science and Technology Council, Taiwan, with Grants No. MOST 111-2917-I-564-005 and NSTC 112-2112-M-003-020-MY3, and the Higher Education Sprout Project of National Taiwan Normal University (NTNU) and the Ministry of Education (MOE) in Taiwan.

\section*{Author contributions}

H.-M.W. and H.-Y.K. contributed equally to this work.
H.-B.C. conceived and conducted the research.
H.-M.W. performed the computations with J.-Y.L.'s help under the supervision of H.-B.C.
H.-M.W. and H.-Y.K. wrote the first draft of the manuscript.
H.-Y.K. and H.-B.C. provided theoretical explanations of the results.

\section*{Competing interests}

The authors declare no competing interests.

\end{document}